# Combining Research Methods and Results from Psychology and Physics in the Investigation of Quantum Mechanics

Douglas M. Snyder

Analysis of modern physical theory indicates that there is a fundamental link between the observing, thinking individual and the physical world (e.g., Epstein, 1945; Renninger, 1960; Snyder, 1986, 1992, 1993, 1994, 1995; Wigner, 1983). This being the case, both psychological research methodology and results should be relevant to the exploration of this link. It is important to provide an example of how psychology can add in a unique way to the exploration of the link between cognition and the physical world found in modern physical theory. A prototype for more complicated investigations is presented through exploring a question in quantum mechanics.

Gedankenexperiments are proposed that indicate that research from psychology is relevant to understanding the relationship between perception and the physical world in quantum mechanics. These gedankenexperiments rely on research in psychology concerning the perceptual and behavioral adaptation of a human observer to inversion of incoming light. The gedankenexperiments presented demonstrate how the uncertainty principle may be affected by psychological phenomena. The uncertainty principle involves the inability in quantum mechanics to simultaneously measure observable quantities described by non-commuting Hermitian operators (e.g., the position and momentum of a particle along a particular spatial axis).[1] Further, the gedankenexperiments show that it may be possible to know more than one value at a time from a set of what appear to be mutually exclusive values for a quantity.

When light is reflected from an object in the world and into the eye, the structure of the eye is such that the image on the retina corresponding to the object is up-down and right-left reversed when compared to the position of the object (Dolezal, 1982; Kandel, Schwartz, & Jessell, 1991). The area of psychological research relevant to the proposed gedankenexperiments specifically concerns the effect of inversion of incoming light on visual experience. (In the context of this paper, unless more precisely specified, the terms "inversion" and "inverted" will refer to either rotation of the incoming light 180° around the line of sight, or the reflection [or flipping] of incoming light between the top and lower halves of the visual field along the horizontal separating them.)  First, this research will be reviewed.  Then the gedanken-experiments will be described.

---

[1] Feynman, Leighton, and Sands's (1965) implication that physical interaction is responsible for the constraints of the uncertainty principle is not the case.

### The Effect of Inversion of Incoming Light
### on Visual Perception and Visually Guided Behavior

In the late 1800s, Stratton (1896, 1897a, 1897b) investigated the effects on visual experience of rotating the incoming light 180° around the line of sight such that the retinal images were right side up instead of being in their customary inverted orientation.  Stratton's results were remarkable.  In commenting on the earlier experiment, he wrote:

> In fact, the difficulty of seeing things upright by means of upright retinal images seems to consist solely in the resistance offered by the long-established experience.  There is certainly no peculiar inherent difficulty arising from the new conditions themselves.  If no previous experience had been stored up to stand in opposition to the new perceptions, it would be absurd to suppose that the visual perceptions in such a case would seem inverted.  Any visual field in which the relations of the seen parts to one another would always correspond to the relations found by touch and muscular movement would give us 'upright' vision, whether the optic image lay upright, inverted, or at any intermediate angle whatever on the retina. (Stratton, 1896, p. 617)

His comments apply as well to the results of the second, more thorough experiment.  In his report on the second experiment, he wrote:

> The inverted position of the retinal image is, therefore, not essential to 'upright vision,' for it is not essential to a harmony between touch and sight, which, in the final analysis, is the real meaning of upright vision.  For some visual objects may be inverted with respect to other visual objects, but the whole system of visual objects can never by itself be either inverted or upright.  It could be inverted or upright only with respect to certain non-visual experiences with which I might compare my visual system--in other words, with respect to my tactual or motor perceptions. (Stratton, 1897b, pp. 475-476)

Subsequent work by other researchers in which all incoming light was rotated 180° around the line of sight (e.g., Ewert, 1930; Snyder & Pronko, 1952) or up-down reversed (the top and bottom halves of the visual field are reversed) (e.g., Dolezal, 1982; Kohler, 1962, 1964) has for the most part provided substantial support for Stratton's finding concerning the relative nature of upright vision. The research indicates that there is a high degree of flexibility of the visual system with regard to inversion of incoming visual stimulation on the retina, including that an observer subject to such reversal quickly regains very significant competency in interacting with the environment. For example, Snyder and Pronko (1952) found in their study:

> During the 30-day period that the inverting lenses were worn, the visuo-motor coordinations were refashioned so that the subject performed even better than before the lenses were put on....Introducing the inverted visual field for 30 days and subsequent "normalization" (lenses removed), [*sic*] modified the learning situation. However, the subject went on learning despite these disrupting factors (p. 116).

In general, visual experience restabilizes quickly considering the relatively very brief period of time that the light is inverted compared to the subjects' life experiences prior to their participation in one of the experiments. Visual experience regains a sense of normalcy and is accompanied by the coordination between touch and vision that Stratton (1897b) wrote is "the real meaning of upright vision" (p. 497) (Dolezal, 1982; Erismann & Kohler, 1953, 1958; Kohler, 1962, 1964; Pronko & Snyder, 1951; Snyder & Pronko, 1952).

As alluded to in the above quote from Snyder and Pronko and as found by Ewert (1930), in the laboratory, competency on sensorimotor tasks developed with unrotated light has been shown to transfer to circumstances where incoming light is rotated 180° (Snyder & Pronko, 1952). Furthermore, increased competency on the same sensorimotor tasks subsequently developed with rotated light has been shown to transfer to circumstances where the incoming light is no longer rotated 180° (Ewert, 1930; Snyder & Pronko, 1952). The learning curve for these tasks was in general fairly smooth, except for a spike when the incoming light was first rotated 180° around the line of sight. In natural settings, individuals wearing an optical apparatus that inverted incoming light have reported such activities as driving an automobile, riding a motorcycle, or riding a bicycle, with a significant degree of skill within a relatively short time of putting on the apparatus for the first time (Dolezal, 1982;

Erismann & Kohler, 1953, 1958; Kohler, 1962; Snyder & Pronko, 1952).  In sum, research has indicated that after a relatively brief period of time exposed to inverted light, visual experience in general appears normal and as this visual experience exists in conjunction with the recaptured competency of the individual in the environment, the visual field is upright, just as it was upright before the incoming light was inverted.  The citations to Erismann and Kohler (1953, 1958) and to Pronko and Snyder (1951) refer to films that provide convincing evidence of both perceptual and behavioral adaptation to inversion of incoming light.

In a related study, Brown (1928) wore goggles with prisms that rotated incoming light $75^{\circ}$ around the line of sight for one week, and he adapted to this rotation to a significant degree.  This adaptation occurred even though he described his apparatus as "too unwieldy" (p. 134) to wear every night on a one-half mile trip to his university where various tests were run.  Other work investigating adaptation of the visual system to alterations in incoming light that did not involve inversion of this light has also indicated a very high degree of flexibility in the operation of the visual system (e.g., Gibson, 1933; Held, 1965; Held & Freedman, 1963).

Stratton (1899) conducted an experiment in which through the use of a specific arrangement of mirrors his body image and immediate surroundings were projected in what without adaptation would be the horizontal direction out from his body at the level just above his shoulders, his head in the reflected image of his body closest to his body and the front of his body facing up.  In the very brief time that Stratton wore the mirror arrangement, he found that visual perception and visually guided action showed significant adaptation to the incoming light.  Stratton wrote:

> The whole experience [in this experiment] was thus so similar to the one with the inverting lenses that I hesitate to present it even in this brief outline.  But under the circumstances the very similarity is a distinct addition to the data from the previous experiment, since it shows that the introduction of the new factor--that of distance--does not prevent an ultimate spatial concord [between the haptic, kinesthetic, and visual senses]....The experiment thus suggests that the principle stated in an earlier paper--that in the end we would feel a thing to be wherever we constantly saw it--can be justified in a wider sense that I then intended it to be taken. (Stratton, 1899, pp. 497-498)

Ewert (1930) and Munn (1955/1965) have disputed the finding that visual experience becomes upright after experience with rotated incoming light. It should be noted, though, that the major concern of Ewert and Munn is not so much the subject's phenomenal experience with rotated light but rather with the interpretation of what this phenomenal experience means. For example, Munn (1955/1965), who was one of Ewert's subjects, wrote:

> Localizing reactions became so automatic at times that a "feeling of normalcy" was present. This is probably the feeling reported by Stratton and interpreted as "seeing right-side up." (p. 293)

Or, Ewert concluded that:

> In all forms of activity where overt localizing responses are present there is rapid adjustment to the distracting visual interference until at the end of 14 days of practice the interference is entirely overcome in some of the activities investigated and almost overcome in the other forms....Constant interference during visual disorientation does not prevent the steady growth of a habit. (Ewert, 1930, pp. 353; 357)

Snyder and Pronko (1952) performed an experiment similar in many respects to Ewert's. Munn (1955/1965) wrote about Snyder and Pronko's work: "The results were essentially like his [Ewert's]" (p. 294). In contrast to Ewert, Snyder and Pronko concluded:

> It appears that perceivings form a behavior sequence going back into the individual's past. If the subject of the present experiment had *always* worn the inverting lenses, his past perceivings would have been of a piece with those of the moment when the question ["Well, how do things look to you? Are they upside-down?" (p. 113)] was directed at him. Obviously, then, they would not have been in contrast with the latter and would not have called attention to themselves. Stated in another way, if this subject had somehow developed amnesia at the point at which he put on the inverting lenses, then things could not appear upside-down because there would be no basis of comparison or contrast. That they did appear upside-down is clearly a strict function of his previously acquired perceivings. (pp. 113-114)

In *A History of Experimental Psychology*, Boring (1929/1950) wrote about Stratton's work:

> In 1896 Stratton put the matter to test, having his subjects [actually only Stratton himself] wear a system of lenses which reversed the retinal image and made it right side up. The expected happened. The perceived world looked upside down for a time and then became reversed. Taking the glasses off resulted once again in reversal which was soon corrected. Stratton was not, however, confused by the homunculus. He described how up was nothing in the visual sensory pattern other than the opposite of down, and that orientation is achieved by the relation of the visual pattern to somothesis and behavior. When you reach up to get an object imaged at the top of the retina, then you have indeed got the visual field reversed and will not find the object unless you have on Stratton's lenses. Ewert repeated this experiment in 1930, *with similar results* [italics added]....Had the view of a freely perceiving agent in the brain not been so strongly entrenched, this problem could not have continued to seem so important in 1604, 1691, 1709, 1838, 1896 and 1930. (p. 678)

Boring knew of Ewert's work and saw that the empirical results obtained by Ewert supported Stratton's conclusion even if Ewert's own conclusion based on the empirical results he found was not in agreement with Stratton's. Dolezal (1982) wrote concerning the results of his experiment and those found in other experiments involving the inversion of incoming light to the observer:

> In the course of living in a world transformed, the observer's initial fears become calmed, he or she finds the discomforts quite tolerable, the strange sights fade and become common, and ineptness changes to competency. (p. 301)

### A Biperceptual Capability

Dolezal proposed that the observer who adapts to inversion of incoming light is *biperceptual* and *biperformatory*. Biperceptual refers to the simultaneous existence of the visual perceptual capabilities associated with both pre-inversion and inversion conditions. These capabilities may be considered distinct reference frames for the individual who has undergone inversion of incoming light. Similarly, biperformatory refers to the simultaneous existence of an individual's capabilities to act competently in the environment both before and after inversion of the incoming light. These capabilities as well are divided into distinct frameworks for the individual who has experienced and adapted to this inversion of incoming light. Dolezal (1982) wrote:

> The adapted observer appears to differ from the unadapted observer in several main respects. After some 200 hours of living with reversing prisms, an observer once again experiences visual stability of the perturbed environment [i.e., up-down reversal of incoming light]. This is true for a wide range of rates of head movements (HMs). Moreover, the adapted observer has acquired what may be called another "personality" (i.e., he or she has the dual facility to be perceptually and emotionally comfortable and to act competently both with and without transforming prisms). The adapted observer is thus a very different creature from the unadapted observer--somewhat like someone with a second language or a novel set of skills that can only be directly displayed under special circumstances (cf. state-dependent learning and recall). The observer becomes what I call biperceptual and biperformatory....In general, the adapted observer is capable of living in both worlds, under both sets of information conditions and behavioral requirements with roughly equal comfort and competence. (p. 297)

Consider the following observation reported by the subject in Snyder and Pronko's study, who happened to be Snyder, that supports Dolezal's thesis:

> Toward the end of the experiment [i.e., the period in which the subject wore the inverting glasses], the subject was adequately adjusted [adapted]. The following insightful experience occurred. He was observing the scene from a tall building. Suddenly someone asked, "Well, how do things look to you? Are they upside-down?"
>
> The subject replied, "I wish you hadn't asked me. Things were all right until you popped the question at me. Now, when I recall how they *did* look *before* I put on these lenses, I must answer that they do look upside down now. But until the moment that you asked me I was absolutely unaware of it and hadn't given a thought to the question of whether things were right-side-up or upside-down." (Snyder & Pronko, 1952, p. 113)

In a study of retention of the effects of such inversion, Snyder and Snyder (1957) found that when the inverted conditions were reintroduced for a subject some time after the subject's initial experience with inverted light, the subject's adjustment the second time to the inverted light indicated that learning occurred as a result of the first experience and had been retained over a two-year period between the first and second experiences with inversion of the incoming light. Specifically, Snyder and Snyder found that the time to complete various tasks consistently took less time in the second experience than in the first. The learning curves in the first and second experiences were very similar for each of the tasks, only in the second exposure the times to complete the tasks were consistently lower than the times to complete them in the first exposure.

In his research, Stratton noted how quickly the perceptual framework of the subject exposed to inverted incoming light could switch between the unadapted and the adapted orientation. He also noted the possibility of their coexistence. For example, on the seventh day of wearing his optical apparatus in his second experiment with inverted light, Stratton (1897b) wrote:

When I watched one of my limbs in motion, no involuntary suggestion arose that it was in any other place or moved in any other direction than as sight actually reported it, except that in moving my arm a slightly discordant group of sensations came from my unseen shoulder. If, while looking at the member, I summoned an image of it in its old position, then I could feel the limb there too. But this latter was a relatively weak affair, and cost effort. When I looked away from it, however, I involuntarily felt it in its pre-experimental position, although at the same time conscious of a solicitation to feel it in its new position. This representation of the moving part in terms of the new vision waxed and waned in strength, so that it was sometimes more vivid than the old, and sometimes even completely overshadowed it. (p. 465)

It is remarkable that the visual system has demonstrated a great degree of flexibility in the inversion experiments given the degree of artificiality introduced into the experimental circumstances by the optical apparatuses that have been used. For example, in the inverted light experiments, Stratton used a device that allowed for incoming light to only one eye while the other eye was covered. Ewert's device was lightweight but allowed for a limited visual field. In an attempt to widen the visual field over that of most other experiments in which all incoming light is inverted for an extended period of time, Dolezal (1982) built his optical device out of a football helmet and inserted glass prisms in the limited space usually left open for a football player to see. His device weighed 8 pounds, 6 ounces. There is further work to be done concerning the effect of inverted light on visual experience and visually guided action. But the basic result that there is significant adaptation in visual experience and visually guided action to inversion of incoming light has been established (Snyder, 1992, 1993, 1995).

*Hard Wiring of the Visual System*
*and the Isotropy of Space*

Originally, Stratton was concerned with showing that two theories concerning inversion of incoming light were incorrect. Essentially, these theories maintained some sort of hard-wiring of either the neural component of the visual system (the projection theory) or its supporting musculature (the eye movement theory). In the projection theory, inversion of the retinal image was needed because of the crossing of the lines of direction of light from external objects when light from the external world moves through the eye. Perception was considered to depend on these lines of direction that projected outward to the upright objects in the physical world from which the light rays originated.

The eye movement theory related to the use of the musculature about the eye to provide definitive information about the correct position of objects in the world. Thus, if the eyes move upward in their sockets, they see the upper parts of objects in the physical world, and if the eyes move downward in their sockets, they see the lower part of these objects in the physical world. In this process, though, movement of the eye upward, for example, results in the lower portion of the retina receiving more of the incoming light. Inversion of the incoming light would help to correct this problem and could provide the basis for indicating the upright nature of the physical world.

Many scholars have believed the tenet behind the projection and eye movement theories that there is only one way the visual system can function in order to perceive the physical world as upright. This is an assumption that Boring (1929/1950) maintained was based on the notion of the homunculus. It can be seen in the descriptions of the projection and eye movement theories that they both have another more basic tenet as an assumption regarding the physical world. This assumption is that the physical world itself has an absolute status regarding its being upright. For example, if the physical world were indeed upside down, would scholars seriously entertain a theory of visual perception based on the hard wiring of the visual system? The assumption that the physical world has an absolute status regarding its being upright violates the isotropy of space in that space is not fundamentally the same in different directions.

### Manipulating Visual Perception in the
### Investigation of a Physical System

How does this adaptability of visual experience and visually guided behavior to inversion of incoming light help in the analysis of modern physical theory? Consider the spin angular momentum of an electron, a quantity intrinsic to the electron. (A classical analog to this quantity would be the angular momentum of a spinning top.) Quantum mechanics predicts that it is possible to measure the component of this momentum along any one of three orthogonal axes, $x$, $y$, and $z$ (three spatial axes all at right angles to one another). The spin component of the electron along the $z$ axis, or any other spatial axis, can have one of two values when measured (i.e., $+1/2\ h/2\pi$ and $-1/2\ h/2\pi$). Assume in an experiment that this measurement of the electron occurs in the following way. Through the use of a Stern-Gerlach apparatus (Eisberg and Resnick, 1974/1985; Liboff, 1993), a nonuniform magnetic field oriented along the $z$ axis is placed in the path of the electron. The value of the spin component of an electron along the $z$ axis can be determined with the Stern-Gerlach apparatus because the electron will take one path in the apparatus if it has one value and will take another path in the apparatus if it has the other value. As an electron exits the apparatus from one of two locations, indicating which one of two paths the electron took through the apparatus and thus the value of its spin component along the axis of the magnetic field of the apparatus, detectors placed at the exits of the apparatus detect the electron. A human observer inspecting these detectors at the exits of the Stern-Gerlach apparatus then can determine which path the electron took and the value of its spin component along the axis of the magnetic field of the apparatus and its gradient.

Assume the $z$ axis is in the vertical direction relative to the subject, appearing to go up and down. The two values of the spin component along the $z$ axis can be designated spin up or spin down according to the motion of electrons passing through the Stern-Gerlach apparatus relative to the vertical axis. Assume that the $y$ axis runs perpendicular to the ideal plane formed by the subject's face and that prior to entering the nonuniform magnetic field the electron is traveling along this axis. Assume that the $x$ axis runs horizontally relative to the subject, from side to side. The experimental circumstances are depicted in Figure 31, where + and - refer to the positive and negative directions along a spatial axis. According to quantum mechanics, precise knowledge resulting from measurement of one of the spin components means that knowledge of each of the other two spin components is completely uncertain. Precise knowledge of the component along the $z$ axis, for example, means that knowledge of each of the components along the $x$ and $y$ axes is completely uncertain.

This limitation concerning the knowledge of certain paired quantities in quantum mechanics also characterizes the simultaneous precise determination of the position and momentum of an electron along a spatial axis (Eisberg and Resnick, 1985; Liboff, 1992). Precise knowledge of the electron's momentum entails complete uncertainty regarding its position, and precise knowledge of its position entails complete uncertainty regarding its momentum. As Einstein, Podolsky, and Rosen (1935) wrote:

> It is shown in quantum mechanics that, if the operators corresponding to two physical quantities, say $A$ and $B$, do not commute, that is, if $AB \neq BA$, then the precise knowledge of one of them precludes such a knowledge of the other. Furthermore, any attempt to determine the latter experimentally will alter the state of the system in such a way as to destroy the knowledge of the first. (p. 778)

Where A and B are the operators corresponding to the components of the position and momentum of an electron, respectively, along a particular spatial axis, they do not commute. Similarly, where A and B are the operators for any two components along orthogonal spatial axes of the spin angular momentum of an electron, they do not commute.

One can use an apparatus like that developed by Stratton to rotate the incoming light to an experimental subject such that the light is rotated around the

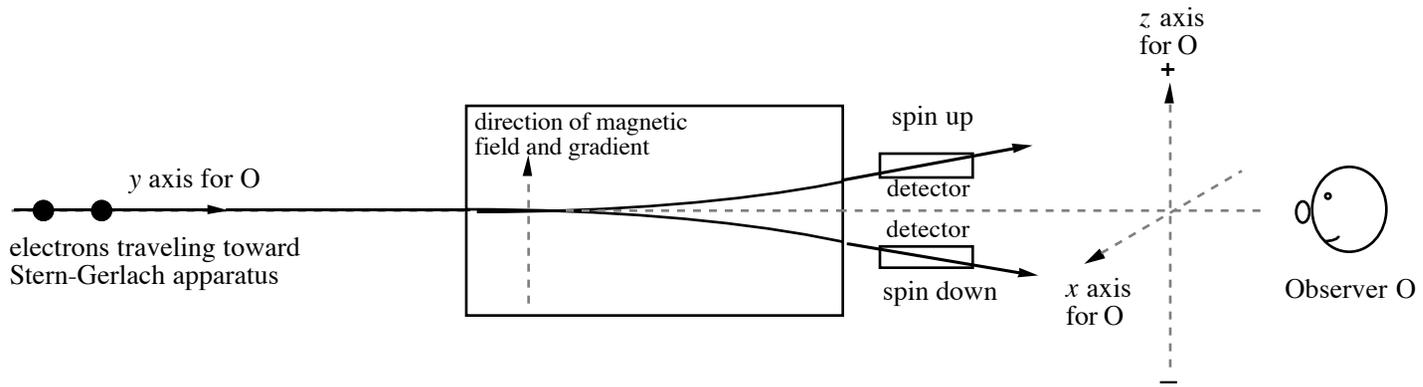

Figure 31

Observer O, not wearing the optical apparatus, viewing the spin component along the
$z$ axis in the spatial structure of O of electrons traveling through a Stern-Gerlach
apparatus without rotation of the visual field.

*y* axis 90$^{\circ}$. Then, what was information concerning the *z* axis is now information concerning the *x* axis as concerns the light impinging on the subject's retina. Given Stratton's results, there is a good possibility that, after a period of orientation with this apparatus, particularly if it is worn for an uninterrupted period and the subject is allowed to move freely in the natural environment, the subject will see the electron moving up or down and not sideways.

Indeed, the natural scenarios tested by Stratton and others are much more complex than the scenario need be that could be presented to an observer in a laboratory setting observing the path of an electron along a spatial axis in an inhomogeneous magnetic field like that created by a Stern-Gerlach apparatus. It does appear that adaptation to inverted light depends to a significant degree on a subject's experience in moving in his or her natural environment while incoming light is inverted by the optical apparatus. Because of the uncomplicated nature of the adaptation needed in the proposed experiment, though, the amount of sensorimotor experience in the natural environment required by the subject for the necessary degree of adaptation of the visual system to occur should not be great. It should be emphasized that research has indicated that the shift in the spatial structure of the observer that depends on sensorimotor experience involves variables pertaining to the individual, for example the correspondence of kinesthetic and haptic sensation with visual perception that develops over time. This correspondence among sensory modalities is not determined by the object or objects in the physical world that are perceived.

Once there is a significant degree of adaptation in the subject's visual experience, according to the information impinging on the subject's retina, the subject is measuring what in the original situation without rotation of the incoming light is the *x* axis (Figure 32). What for observers in the original situation is up and down along the *z* axis is for the subject up and down along the *x* axis, where the *x* axis is labeled in terms of the spatial structure associated with the observer who does not wear the optical apparatus. For the subject wearing the optical apparatus and adapted to the rotated incoming light, the spin components along the *y* and *z* axes of the spatial structure associated with the observer who does not wear the optical apparatus are completely uncertain.

It is important to note that for the observer wearing the optical apparatus and adapted to the rotated light, in his or her shifted spatial structure, the *z* axis runs along the vertical or upright direction just as the *z* axis runs along the

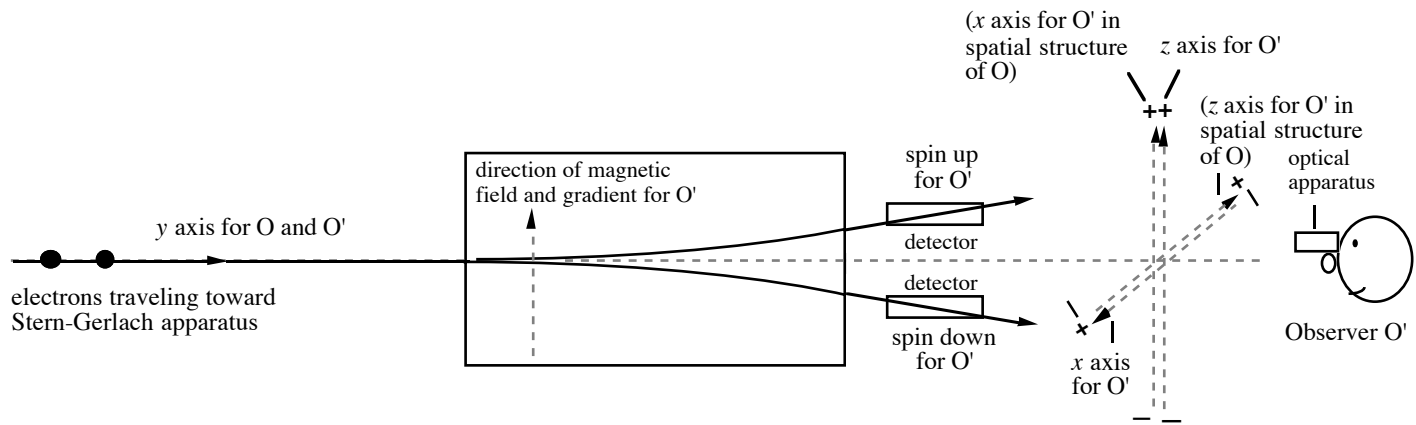

Figure 32

Observer O', adapted to wearing the optical apparatus, viewing the spin component along the $z$ axis in the spatial structure of O' of electrons traveling through a Stern-Gerlach apparatus with 90 degree rotation of the visual field, in terms of the spatial structure of O, around the $y$ axis. ($x$ and $z$ axes for O' in spatial structure of O, who does not wear the optical apparatus, are shown.)

vertical or upright direction for the observer who does not wear the optical apparatus. The observer adapted to the incoming light and wearing the optical apparatus measures the spin component in his or her shifted spatial structure along the $z$ axis, just as the observer who does not wear the optical apparatus measures the spin component along the $z$ axis in his or her spatial structure.

It appears possible to know the precise values for two components along orthogonal axes of the spin angular momentum of an electron at the same time for a particular spatial structure, where one takes into account one of the orthogonal axes as it is perceived by an observer using a different spatial structure. The key to this gedankenexperiment is that a basic shift has occurred in the spatial structure for the observer who wears the optical apparatus and adapts to the rotated incoming light. This shifted spatial structure has the same significance for this observer as the spatial structure of the observer who does not wear the optical apparatus has for this latter observer. This basic gedankenexperiment shows in a particular instance that it is possible that the uncertainty principle is conditioned by a link between the spatial structure of an observer in the physical world and the orientation of incoming light relative to the observer. Of particular importance is the evidence that there are factors specific to the observer in the development of an individual's spatial structure.

Evidence supporting a biperceptual character of visual perception after adaptation to inversion of incoming light has been noted. This biperceptual character of visual perception concerns the simultaneous existence of the distinct visual perceptual capabilities associated with both pre-inversion and post-inversion conditions. It may be possible for one subject in the experiment outlined above involving spin angular momentum components along orthogonal spatial axes to be involved in mutually exclusive situations simultaneously concerning the same concrete physical circumstances. That is, the adapted subject may be able to instantly shift from being involved in one of the experimental scenarios to the other.

*A 180° Rotation*

There is another circumstance besides that already discussed that is perhaps even more surprising. Consider that an optical apparatus is used that rotates incoming light $180°$ around the line of sight (Figure 33). For the subject wearing the apparatus but not yet adapted, the negative direction of the $z$ axis is associated with spin up and the positive direction of the $z$ axis is associated with spin down. Once a significant degree of adaptation in the subject's visual experience occurs, when the subject observes that an electron has spin up in the positive direction of the $z$ axis, *according to the information impinging on the subject's retina* the subject is measuring what in the original situation without rotation of the incoming light is spin up in the negative direction of the $z$ axis. Similarly, when the subject observes that an electron has spin down in the negative direction of the $z$ axis, according to the information impinging on the subject's retina the subject is measuring what in the original situation without rotation of the incoming light is spin down in the positive direction of the $z$ axis.

If this result is considered in terms of the Schrödinger cat gedankenexperiment (Schrödinger, 1935/1983), it is as if in one situation, one atom of the radioactive material decayed leading to the cat's being dead when observed. Simultaneously, in the other situation, none of the radioactive material decayed, leading to the cat being alive when observed. One situation involves the observer who has not worn the optical device and who is not wearing the device when he or she observes the electron. The other situation involves the observer who is wearing the apparatus when observing the electron and who has adapted to the rotated visual field. In contrast to the Schrödinger cat gedankenexperiment, each observer's perception is similar in structure. But while the effect of the rotation of incoming light for a subject is mitigated upon adaptation, the historical physical event of the rotation of the incoming light due to the optical apparatus remains for the observer wearing the device and acts to distinguish the two situations of the observers.

Practically speaking, the "righting" of the visual field for the adapted subject means that the electron with spin up in the original spatial structure now has spin down in the original spatial structure even though the electron has spin up in the spatial structure of the adapted observer. For the subject wearing the optical apparatus but who has not adapted to the inversion of incoming light, the spatial structure of the world is the same as that for the subject who does not wear the optical apparatus and for whom light is not inverted. The unadapted subject knows that his visual experience is altered by wearing the optical apparatus but that the world itself remains unchanged. The unadapted subject knows that the spin component of the electron is up, that is up for an observer not wearing the optical apparatus, though it appears to be down to him, opposite to the direction found by an observer not wearing the optical apparatus.

But with adaptation to inverted incoming light, including the sense of normalcy in visual experience and competency in visually guided action that

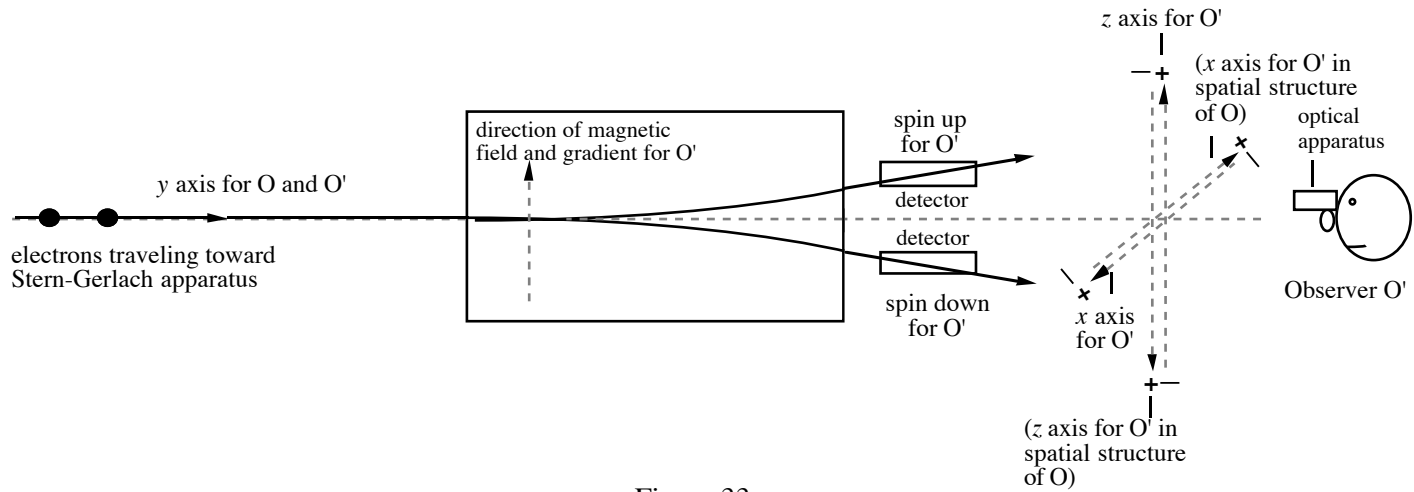

Figure 33

Observer O', adapted to wearing the optical apparatus, viewing the spin component along the *z* axis in the spatial structure of O' of electrons traveling through a Stern-Gerlach apparatus with 180 degree rotation of incoming light, in terms of the spatial structure of O, around the *y* axis. (*x* and *z* axes for O' in spatial structure of O, who does not wear the optical apparatus, are shown.)

supports this sense of normalcy, what appears as a spin up component for the electron to the adapted observer is a spin down component when considered in terms of the original spatial structure, the spatial structure of the subject for whom incoming light is not inverted. Unlike the unadapted observer, the adapted observer does not maintain that the world is unchanged while his visual experience has been altered by inverting incoming light with the use of the optical apparatus. The adapted observer experiences the "real" direction of the spin component opposite to that experienced by the unadapted observer. In terms of the impact of incoming light on his retina, the adapted observer who observes a spin up component *experiences* a spin component in the opposite direction to that which the same pattern of light on the retina supports for the unadapted observer (Snyder, 1995).

*A Possible Objection*

One might object to this conclusion regarding the "righting" of the visual field for the adapted subject in the following way. The artificial reorientation of the incoming light does not prevent tracing back light impacting the retina to the physical existent that is the source of the light so that the actual spin component of the electron along a particular axis in space and the visual perceptions of the electron's spin component in space by the observer for whom light is not inverted, or otherwise rotated, and by the individual adapted to the incoming, altered light are all in agreement. That is, both observers correctly deduce the spin component of the electron because they correctly perceive the motion of the electron in space along the spatial axis along which it is traveling. Then the visual system could be said to simply adapt to the artificial change in incoming light but that the internal sensory coordination by the observer ultimately reflects the absolute positioning of the electron's motion in space as it moves through the Stern-Gerlach device.

But, consider that instead of using an external optical system to alter the orientation of incoming light, another method is used in the gedanken-experiment to change the orientation of light on the retina. Consider that the retinas of an observer with their supporting physiological structures in the eye, in particular the optic nerve, are rotated $180^{\circ}$. In this arrangement, an optical apparatus external to the visual system is no longer used. (Both procedures accomplish the same goal, rotation of incoming light $180^{\circ}$ on the retina.) The results, though, of using either the external optical system or reorienting the retinas themselves should be the same. The human subject whose retinas have been reoriented should with adaptation see the world upright even though the incoming light falls on the retinas in the opposite manner to that found for the observer whose retinas have not been reoriented, where space is considered in terms of the spatial structure of this latter observer. By reorienting the retinas themselves, one can no longer reasonably subscribe to the thesis that the visual system is simply accounting for the artificial nature of the external optical apparatus in order to correctly ascertain the absolute positioning of the electron's motion in space. Here, no extra instrument is added. Only the orientation of the incoming light relative to the retina has been changed.

## Summary of the Results
## of the Gedankenexperiments

Psychology can add in a unique way to the exploration of the link between cognition and the physical world found in modern physical theory. Developing an example of an experiment that combines methodology from both psychology and physics has provided an interesting result. The gedanken-experiments presented have indicated the importance of an individual's spatial structure in measurement in quantum mechanics. Two fundamental features of quantum mechanics are:

1) quantities represented by non-commuting Hermitian operators cannot be simultaneously known;

2) only one value from a set of mutually exclusive values for a quantity can be known at a time.

The gedankenexperiments indicate that both of these features are conditioned by a link between the spatial structure of an observer in the physical world and the orientation of incoming light relative to the observer. An individual's spatial structure itself is dependent on variables pertaining to the observer, for example, the correspondence of the haptic, kinesthetic, and visual senses. Within a particular spatial structure, the two fundamental features of quantum mechanics are correct. Yet, when different spatial structures of different observers, or different spatial structures of the same individual who is adapted to wearing the optical apparatus, are compared, it is likely that these features can be modified. Instead, in these latter circumstances, it is possible that:

1) quantities represented by non-commuting Hermitian operators can be simultaneously known when these quantities are considered in different spatial structures;

2)     more than one value for a quantity from a set of what would be in one spatial structure mutually exclusive values can be known at a time.

The model for developing the gedankenexperiments presented is based on adjusting the observer variables affecting the act of perception. The data obtained in the measurement process concern quantities of the measured physical system, and the expected empirical results are those predicted in quantum mechanics. This model contrasts with most research in quantum mechanics in which the focus is on adjusting variables directly affecting the physical system measured or a physical apparatus directly acting as a measuring instrument.